\RequirePackage{fix-cm}
\documentclass[smallextended]{svjour3}       
\smartqed  
\usepackage{graphicx}
\usepackage{multirow}

\begin{document}

\title{Isobaric heat capacity, isothermal compressibility and fluctuational properties of 1-bromoalkanes}


\titlerunning{Isobaric heat capacity...}        

\author{V.I. Korotkovskii \and O.S. Ryshkova \and Yu.A. Neruchev \and 
        A.L. Goncharov \and E.B. Postnikov\footnote{Corresponding author}}

\institute{
V.I. Korotkovskii, O.S. Ryshkova, Yu.A. Neruchev \at 
							Laboratory of Molecular Acoustics, Kursk State University, Radishcheva st., 33,  Kursk 305000, Russia \\
A.L. Goncharov, E.B. Postnikov \at
              Department of Theoretical Physics, Kursk State University, Radishcheva st., 33,  Kursk 305000, Russia \\
              Tel.: +7-4712-51-04-69\\
              Fax: +7-4712-51-04-69\\
              \email{postnicov@gmail.com}           
           \and
           }

\date{Received: date / Accepted: date}

\maketitle

\begin{abstract}
We present results of the experimental measurements of the isobaric heat capacity  for  1-bromohexane, 1-bromoheptane, 1-bromooctane, 1-bromononane, 1-bromodecane, 1-bromoundecane, 1-bromododecane and 1-bromo-tetradecane at normal pressure and the speed of sound and the density for 1-bromotetradecane within the temperature range 298.15--423.15~K. These data on the isobaric heat capacity and the literature-based reference data  for the density and the speed of sound were used to calculate the isothermal compressibility and the inverse reduced fluctuations. Based on the comparison of the results for pure n-alkanes and $\alpha,\omega$-dibromoalkanes, we discuss the influence of bromine atom on the volume fluctuations.

\end{abstract}

\section{Introduction}

The molecular acoustics approach has been known for a long time as a tool, which allows the  determination of details of intermolecular interactions using experimental data \cite{MathesonBook}, which can be obtained relatively simple. Modern advances in the ultrasound technique, see for review, e.g. \cite{Kaatze2008}, supplied with the thermophysical studies \cite{WeirBook} allow the high-accurate characterization of microscopic properties of the liquids, their phase equilibria and transitions. 

Recently, it has been shown \cite{Goncharov2013} that the knowledge of the speed of sound $c$ and the heat capacity ratio $\gamma$ (or the isothermal compressibility and the density) provides an opportunity to explore structural properties of normal liquids based on the inverse reduced fluctuations
\begin{equation}
\nu=\frac{\mu_0 c^2}{\gamma RT},
\label{nu}
\end{equation}
where $\mu_0$, $R$,and  $T$ are the molar mass, the gas constant and the temperature, respectively. 

The function (\ref{nu}) is the inverse ratio of the relative volume fluctuation to its value in the hypothetical case where the substance acts as an ideal gas for the same temperature--volume parameters. Thus, the absence of intermolecular iterations results in the value $\nu=1$. Therefore, a deviation of this quantity from unity characterizes intermolecular interactions originated from the both coupled force and geometric packing properties. 

The principal goal of the present work is to study an effect of the halogen atom insertion into n-alkanes on the mentioned parameter. This class of substances is quite demonstrative for such an analysis due to the presence of extensive studies of linear hydrocarbons, which show that their thermodynamic properties can be effectively modelled considering the Flory-like approach of almost uniformly distributed subunits (individual atoms or molecule's segments) \cite{Martin1998,Cerdeirina2004,Neruchev2008}. An atom of halogen violates this symmetry that should be apparently exhibited in the fluctuation analysis.

Recently, a variety of data have been obtained for the series of consequent lengths in chains of bromo-substituted alkanes and different numbers of bromine atoms per molecule \cite{Ernst2000,Bolotnikov2007,Ryshkova2009,Bolotnikov2009,Chorazewski2010,Chorazewski2015}. Some gaps in these datasets are filled and presented in the article.  Therefore, these substances could be chosen as an example.

\section{Experimental Part}

\subsection{Materials}

The studied liquids were obtained from Acros Organics (1-bromohexane with an initial purity $>99\%$ in mass; 1-bromoheptane, 99~\%; 1-bromooctane, 99~\%; 1-bromododecane, 98~\%; 1-bromotetradecane, 98~\%), Sigma-Aldrich (1-bromononane, 98~\%), Fluka (1-bromodecane, 97~\%, 1-bromoundecane, 97~\%) and  were used without an additional purification. The purity was controlled via the comparison of the density and the refractive index before and after experiments. The purity of 1-bromotetradecane, taken as an example of samples, was additionally controlled by chromatography using the Agilent 6850 system with the 5973N mass selective detector by Agilent Technologies. The measurements of the peak intensity corresponding to 1-bromotetradecane with regard to other substances present  show that the purity of the used 1-bromotetradecane sample exceeds 98.5~\%.

\subsection{Measurements}

The measurements of the isobaric heat capacity were evaluated at the normal atmospheric pressure conditions within the temperature interval 298.15--423.15~К using the modified differential scanning calorimeter with the monotonic heating \cite{GSSD2}. 

A studied sample was placed into the $1\,\mathrm{cm}^3$ thin-walled cylindrical metal chamber supplied with the thin metallic tube (the cross section area is  $3\,\mathrm{mm}^2$), which ensures that the evaporation occurs sufficiently out of the working zone and, therefore, can be considered as negligible. In addition, it provides the practically constant volume of the liquid within the main chamber preventing its deformation. The main part of the heat flux (through the bottom face) was inspected by the temperature difference $\Delta T$. The temperature was measured by the thermocouple, and the differential temperature was registered by a computer with the frequency 135 readouts per 100~s. A small part of the heat flux received through the lateral surface and determined by the heat conductivity of the liquid was controlled by the calibration procedure.

The resulting procedure was reduced to the measurement of the delay time defined as $\Delta t=\Delta T/d\tau/dt$ utilizing the standard technique of the  scanning calorimetry \cite{HoenneBook2013}: the heating rate $d\tau/dt$ and the temperature difference $\Delta T$ were used as principal measured quantities for the studied liquid and the reference liquid.  N-dodecane \cite{NIST} was used as the reference liquid. 

The desired values of the heat capacity were calculated as 
$$
C_p=C_p^{ref}\frac{\rho_{ref}}{\rho}\frac{(\Delta t+\Delta t')}{(\Delta t_{ref}+\Delta t'')}.
$$
Here $C_p$, $C_p^{ref}$, $\rho$, $\rho_{ref}$ are the isobaric heat capacities and the densities of the studied and the reference liquids, correspondingly; $\Delta t$, $\Delta t_{ref}$ are the delay times of the heat flow meter for the studied and the reference liquids ; $\Delta t'$ and $\Delta t''$ are proportional to the heat conductivities of both liquids and  determined by a calibration of the setup. 

The resulting relative errors of measurements (2\%) are determined by the error of differential temperature measurements (1\%), of its heating rate (0.3\%), of the mass and density of a sample (0.1\%) and of the calibration using a reference liquid (0.5\%). 

Due to the absence of known open data for the speed of sound and the density of 1-bromotetradecane within the required range of thermodynamic conditions, the corresponding measurements were evaluated.

The speed of sound was measured along the coexistence line within the temperature interval  293.15--423.15~К by the pulse phase method \cite{Neruchev2005} within the dispersionless region (the frequency is equal to 1~MHz) using the experimental device developed in the Laboratory of Molecular Acoustics (Research Center for Condensed Matter Physics, Kursk State University) and certified by the Russian State Service for Standard Reference Data \cite{GSSD}. 

The temperature was kept constant with the relative error $0.01$~K using the thermostat ``Termex'' (Russia). Liquid's temperature was controlled and measured using the platinum resistance thermometer (100~Ohm) placed within the measurement chamber connected with the signal converter ``Tercon'' by ``Termex'' (Russia) with the accuracy within $\pm0.02$~K. The general estimation of all factors affecting the speed of sound provides the value relative errors with an upper limit $0.1\%$.

The density measurements were evaluated using a quartz pycnometer at the normal atmospheric pressure within the temperature interval 263.15--423.15~K. The pycnometer was calibrated using density's data for water, n-heptane and n-dodecane \cite{NIST}. The corrections with respect to the air's buoyancy force and pycnometer's thermal expansion reduce the relative errors of density measurements to $0.01\%$. 

\subsection{Results}

The raw data for the speed of sound and the density of 1-bromotetradecane are presented in Table~1.

\begin{table}
\label{rawc}
\caption{The results of measurements of the density and the speed of sound for 1-bromotetradecane}
\begin{tabular}{c|c|c}
\hline
$T\, (\mathrm{K})$&$c\, (\mathrm{m\cdot s^{-1}})$&$\rho\, (\mathrm{kg\cdot m^{-3}})$\\
\hline
298.15&1277.2&1012.2\\
303.15&1260.6&1008.1\\
308.15&1244.1&1003.9\\
313.15&1227.9&999.8\\
318.15&1211.9&995.7\\
323.15&1196.0&991.6\\
328.15&1180.4&987.4\\
333.15&1164.9&983.3\\
338.15&1149.6&979.1\\
343.15&1134.5&974.9\\
348.15&1119.5&970.8\\
353.15&1104.6&966.6\\
358.15&1089.9&962.4\\
363.15&1075.3&958.2\\
368.15&1060.9&954.0\\
373.15&1046.5&949.8\\
378.15&1032.3&945.6\\
383.15&1018.2&941.3\\
388.15&1004.2&937.1\\
393.15&990.2&932.9\\
398.15&976.4&928.6\\
403.15&962.6&924.4\\
408.15&948.9&920.1\\
413.15&935.3&915.9\\
418.15&921.7&911.6\\
423.15&908.1&907.4\\
\hline

\end{tabular}

\end{table}

These data and the reprocessed data on the speed of sound and the density given in \cite{Bolotnikov2009,Bolotnikov2007} were approximated by the quadratic $c=A_0+A_1T+A_2T^2$ and the cubic $\rho=A_0+A_1T+A_2T^2+A_2T^3$ polynomials correspondingly. The coefficients are given in Tables~2-3.

The results of the measurements of the isobaric heat capacity are listed in Table~4.

\begin{figure}
\includegraphics[width=\textwidth]{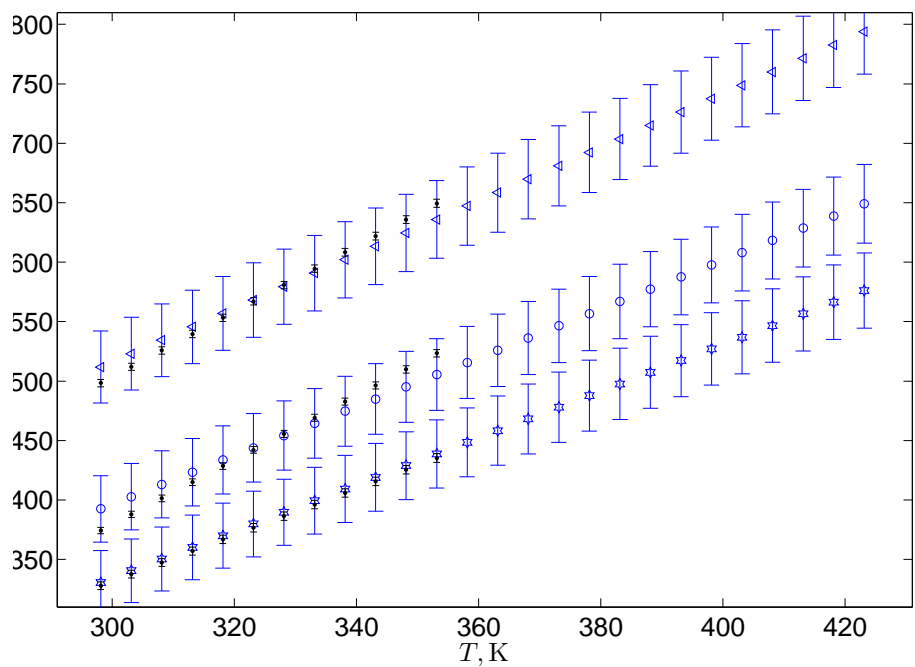}
\caption{The comparison of the measured values of the isobaric heat capacity for 1-bromohexane (hexagons), 1-bromoheptane (circles), and 1-bromodecane (triangles) with the data of \cite{Chorazewski2005} marked by black dots  supplied by the error bars accordingly to the information provided in the cited work}.
\label{figcompar}
\end{figure}

Fig.~\ref{figcompar} shows the comparison of our results with the data \cite{Chorazewski2005} obtained in a more narrow interval of temperatures. They practically coincide around room temperature; deviation increases with the growth of the temperature. However, it does not  exceed the averaged error of measurements.

\begin{figure}
\includegraphics[width=\textwidth]{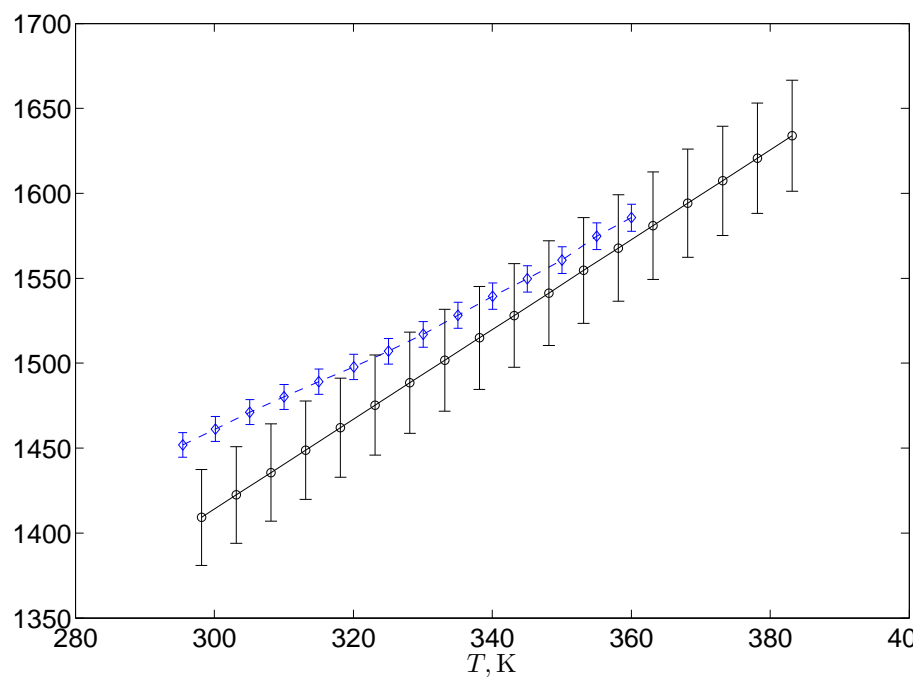}
\caption{The comparison of the measured values of the isobaric heat capacity for 1-bromooctane (circles) with the known data for the saturated heat capacity \cite{Becker2000}(diamonds).}
\label{figcompar2}
\end{figure}

Fig.~\ref{figcompar2} shows the comparison of our results with the data for the saturated heat capacity\cite{Becker2000} for 1-bromooctane. Note that the temperature interval is taken below the boiling point ($T_b=474\,K$), i.e., a relative difference between the isobaric and the saturated heat capacities (usually around $0.1\%$) should not exceed the relative error of measurements ($2\%$ and $0.5\%$ correspondingly). One can see that both data ranges overlap for $T\geq 320\,K$ and are close for smaller temperatures. On the other hand, one can detect a sudden deviation of points to larger values in the data of Becker et al. for $T<320\,K$, i.e. there are questions about their accuracy in this region since there is no such effect for other liquids of a similar molecular structure, see Fig.~\ref{figcompar}.

\begin{table}
\label{Ac}
\caption{The coefficients of the polynomial approximating the speed of sound.}

\begin{tabular}{l|ccc}
\hline
&$A_0\,(\mathrm{m\cdot s^{-1}})$&$A_1\,(\mathrm{m\cdot s^{-1}\cdot K^{-1}})$&$A_2\cdot10^3\,(\mathrm{m\cdot s^{-1}\cdot K^{-2}})$\\
\hline
1-bromohexane&	2223.9&	-4.34065&	1.8065\\
1-bromoheptane&	2260.74&	-4.37093&	1.8860\\
1-bromooctane&	2293.64&	-4.40209&	1.1931\\
1-bromononane&	2323.06&	-4.42538&	1.9983\\
1-bromodecane&	2351.06&	-4.47973&	2.1122\\
1-bromoundecane&	2375.13&	-4.48107&	2.0980\\
1-bromododecane&	2414.13&	-4.62387&	2.3189\\
1-bromotetradecane&	2478.20&	-4.79997&	2.5794\\
\hline
\end{tabular}

\end{table}

\begin{table}
\label{Arho}
\caption{The coefficients of the polynomial approximating the density.}

\begin{tabular}{l|cccc}
\hline
&$A_0$&$A_1$&$A_2\cdot10^3$&$A_3\cdot10^6$\\
&$(\mathrm{kg\cdot m^{-3}})$&$(\mathrm{kg\cdot m^{-3}\cdot K^{-1}})$&$(\mathrm{kg\cdot m^{-3}\cdot K^{-2}})$&$(\mathrm{kg\cdot m^{-3}\cdot K^{-3}})$\\
\hline
1-bromohexane&1504.09&-1.1780&0.3842&0.7716\\
1-bromoheptane&1490.15&-1.5892&2.1526&-2.8206\\
1-bromooctane&1392.80&-0.9281&-0.070002&-0.14485\\
1-bromononane&1409.08&-1.4308&1.7356&-2.0668\\
1-bromodecane&1370.39&-1.2595&1.1955&-1.3834\\
1-bromoundecane&1383.82&-1.6233&2.4011&-2.5700\\
1-bromododecane&1376.72&-1.6890&2.6883&-2.8971\\
1-bromotetradecane&1245.40&-0.7426&-0.13268&0.0000001\\
\end{tabular}
\end{table}

\begin{table}
\label{Cptab}
\caption{The measured values of the isobaric heat capacity ($\mathrm{J\cdot kg^{-1}K^{-1}}$) for the series of 1-bromoalkanes.}
\begin{tabular}{c|cccccccc}
\hline 
$T\,(\mathrm{K})$ &$\mathrm{C_6\cdots}$&$\mathrm{C_7\cdots}$&$\mathrm{C_8\cdots}$&
$\mathrm{C_9\cdots}$&$\mathrm{C_{10}\cdots}$&$\mathrm{C_{11}\cdots}$&$\mathrm{C_{12}\cdots}$&$\mathrm{C_{14}\cdots}$\\ 
\hline 
298.15 & 1330.7 & 1392.5 & 1409.2 & 1488.3 & 1511.8 & 1586.8 & 1602.1 & 1625.9 \\ 
303.15& 1340.5&1402.8&1422.4&1499.2&1523.0&1597.4&1614.5&1637.6\\
308.15&1350.3&1413.0&1435.6&1510.1&1534.3&1608.0&1626.9&1649.2\\
313.15&1360.1&1423.3&1448.8&1521.1&1545.6&1618.6&1639.3&1660.9\\
318.15&1369.9&1433.6&1462.0&1532.0&1556.9&1629.2&1651.7&1672.6\\
323.15&1379.8&1443.8&1475.3&1542.9&1568.2&1639.8&1664.1&1684.3\\
328.15&1389.6&1454.1&1488.5&1553.8&1579.5&1650.4&1676.5&1696.0\\
333.15&1399.4&1464.4&1501.7&1564.8&1590.8&1661.0&1688.9&1707.7\\
338.15&1409.2&1474.6&1514.9&1575.7&1602.0&1671.6&1701.3&1719.3\\
343.15&1419.0&1484.9&1528.1&1586.6&1613.3&1682.2&1713.7&1731.0\\
348.15&1428.9&1495.1&1541.3&1597.5&1624.6&1692.8&1726.1&1742.7\\
353.15&1438.7&1505.4&1554.6&1608.5&1635.9&1703.3&1738.4&1754.4\\
358.15&1448.5&1515.7&1567.8&1619.4&1647.2&1713.9&1750.8&1766.1\\
363.15&1458.3&1525.9&1581.0&1630.3&1658.5&1724.5&1763.2&1777.7\\
368.15&1468.1&1536.2&1594.2&1641.2&1669.8&1735.1&1775.6&1789.4\\
373.15&1477.9&1546.5&1607.4&1652.2&1681.0&1745.7&1788.0&1801.1\\
378.15&1487.8&1556.7&1620.7&1663.1&1692.3&1756.3&1800.4&1812.8\\
383.15&1497.6&1567.0&1633.9&1674.0&1703.6&1766.9&1812.8&1824.5\\
388.15&1507.4&1577.2&1647.1&1684.9&1714.9&1777.5&1825.2&1836.2\\
393.15&1517.2&1587.5&1660.3&1695.9&1726.2&1788.1&1837.6&1847.8\\
398.15&1527.0&1597.8&1673.5&1706.8&1737.5&1798.7&1850.0&1859.5\\
403.15&1536.8&1608.0&1686.8&1717.7&1748.8&1809.3&1862.3&1871.2\\
408.15&1546.7&1618.3&1700.0&1728.6&1760.0&1819.8&1874.7&1882.9\\
413.15&1556.5&1628.6&1713.2&1739.6&1771.3&1830.4&1887.1&1894.6\\
418.15&1566.3&1638.8&1726.4&1750.5&1782.6&1841.0&1899.5&1906.2\\
423.15&1576.1&1649.1&1739.6&1761.4&1793.9&1851.6&1911.9&1917.9\\
\hline 
\end{tabular} 
\end{table}

These data were used for the calculation of the isothermal compressibility using the standard thermodynamical equation
$$
\beta_T=\frac{1}{\rho c^2}+\frac{\alpha_p^2T}{\rho C_p},
$$
where the isobaric coefficient of the thermal expansion is calculated by the numerical differentiation of the function interpolating the density at the normal pressure. The resulting values are presented in Table~5.

\begin{table}
\label{betatab}
\caption{The isothermal compressibility  ($\mathrm{10^{-10}\,Pa^{-1}}$) for the series of studied 1-bromoalkanes.}
\begin{tabular}{c|cccccccc}
\hline 
$T\,(\mathrm{K})$ &$\mathrm{C_6\cdots}$&$\mathrm{C_7\cdots}$&$\mathrm{C_8\cdots}$&
$\mathrm{C_9\cdots}$&$\mathrm{C_{10}\cdots}$&$\mathrm{C_{11}\cdots}$&$\mathrm{C_{12}\cdots}$&$\mathrm{C_{14}\cdots}$\\ 
\hline
298.15&9.091&8.642&8.385&8.035&7.872&7.635&7.504&7.251\\
303.15&9.404&8.928&8.652&8.285&8.108&7.860&7.722&7.467\\
308.15&9.731&9.228&8.930&8.546&8.353&8.094&7.948&7.690\\
313.15&10.073&9.541&9.218&8.817&8.607&8.337&8.184&7.920\\
318.15&10.429&9.869&9.518&9.099&8.871&8.589&8.429&8.156\\
323.15&10.802&10.212&9.829&9.394&9.146&8.853&8.683&8.400\\
328.15&11.191&10.571&10.154&9.701&9.431&9.127&8.949&8.652\\
333.15&11.598&10.948&10.491&10.021&9.728&9.412&9.225&8.912\\
338.15&12.024&11.342&10.843&10.355&10.037&9.709&9.513&9.180\\
343.15&12.469&11.756&11.209&10.704&10.358&10.019&9.813&9.457\\
348.15&12.936&12.190&11.591&11.069&10.693&10.343&10.126&9.743\\
353.15&13.425&12.645&11.989&11.449&11.042&10.680&10.452&10.038\\
358.15&13.938&13.123&12.404&11.847&11.405&11.031&10.791&10.344\\
363.15&14.476&13.626&12.837&12.262&11.783&11.398&11.146&10.660\\
368.15&15.041&14.154&13.289&12.697&12.178&11.782&11.516&10.987\\
373.15&15.634&14.709&13.762&13.152&12.589&12.182&11.902&11.326\\
378.15&16.257&15.294&14.255&13.628&13.018&12.600&12.306&11.677\\
383.15&16.912&15.909&14.771&14.126&13.466&13.037&12.727&12.041\\
388.15&17.601&16.558&15.310&14.649&13.934&13.494&13.167&12.419\\
393.15&18.327&17.241&15.875&15.196&14.423&13.972&13.628&12.811\\
398.15&19.092&17.962&16.466&15.770&14.933&14.473&14.109&13.219\\
403.15&19.898&18.722&17.085&16.372&15.467&14.996&14.612&13.642\\
408.15&20.748&19.525&17.734&17.004&16.025&15.545&15.139&14.083\\
413.15&21.646&20.373&18.414&17.667&16.609&16.119&15.690&14.543\\
418.15&22.594&21.270&19.127&18.364&17.220&16.722&16.268&15.022\\
423.15&23.596&22.219&19.876&19.096&17.859&17.353&16.872&15.521\\
\hline 
\end{tabular} 
\end{table}

\section{Inverse reduced fluctuations for 1-dibromoalkanes}

Now we can explore the dependence of the inverse reduced volume fluctuations on the density for the studied bromoalkanes. It is easy to show applying the definition of the speed of sound squared as $c^2=(\rho \beta_S)^{-1}$ and $\gamma=\beta_T/\beta_S$,  Eq.~(\ref{nu}) can be represented as
$$
\nu=\frac{\mu_0}{RT}\frac{1}{\rho\beta_T}.
$$

\begin{figure}
\includegraphics[width=\textwidth]{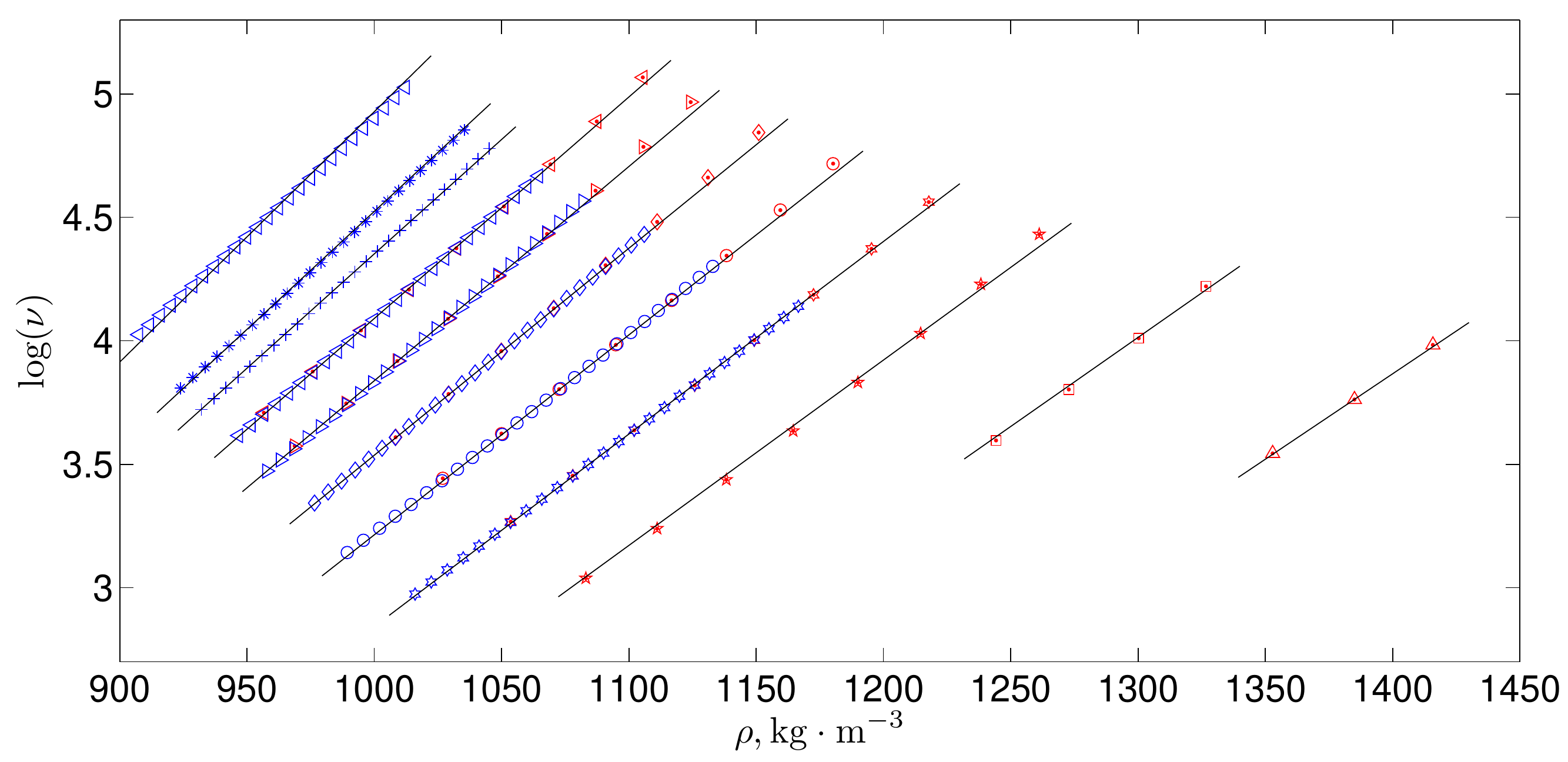}
\caption{The logarithms of the inverse reduced fluctuations as a function of the density for the series of liquids from 1-tetradecane to 1-bromopropane (from left to right). The open markers (blue in color online) corresponds to the values calculated the data presented in this work; the markers with dots inside indicate values obtained using the data from \cite{Ryshkova2009}; solid lines are drawn using the proposed linear fit for slope coefficients of the series.}
\label{Bralks}
\end{figure}

Thus, substituting into this expression the values from Table~5 and the densities calculated using the polynomials with the coefficients given in Table~3, we compute the desired values of $\nu$. The corresponding plots are represented in Fig.~\ref{Bralks} as open markers. One can see that all points fall into the straight lines in semilogarithmic co-ordinates, i.e. the density dependence of the inverse reduced volume fluctuations is exponential one ($\nu=\exp(\kappa\rho+b)$) for the considered range of PVT states. Such a behaviour is expectable. It has been detected before for simple liquids (liquid saturated noble gases \cite{Goncharov2013}) and pure n-alkanes \cite{Postnikov2014}. Besides, the high accuracy of the linear fit supports the conclusions about the accuracy of measurements. 

For the comparison, we show in Fig.~\ref{Bralks} also the results calculated using the data provided in \cite{Ryshkova2009} for a more wide temperature interval (but with a coarser step size) for the series from 1-bromohexane to 1-bromodecane. In addition, these data allow for covering of short-chained 1-bromoalkanes, from 1-bromopropane to 1-bromopentane. It should be pointed out that the results of calculations based on \cite{Ryshkova2009} demonstrate the larger irregular deviations from straight lines and some overestimations of the inverse reduced fluctuations in the region of lowermost temperatures. Thus, the thermophysical data presented in the section above can be argued as more warrant using this indirect sensitive criterion. 

Looking at Fig.~\ref{Bralks}, one can mention that all lines have different slopes $\kappa$ growing with the increasing length of the chain. The corresponding plot is shown in Fig.~\ref{figallslopes}. The point corresponding to the majority substances  satisfies the linear fit 
\begin{equation}
\kappa_n=2.92\cdot 10^{-4}n+6.05\cdot 10^{-4},
\label{1Bfit}
\end{equation}
where $n$ is the number of carbon atoms. 
The sufficient outlier is $\kappa$ for 1-tetradecane.
Such a behaviour will be discussed below. 

The solid lines in Fig.~\ref{Bralks} are drawn by applying this analytical approximation for $\kappa$ and $b$ determined via the least mean square deviation of the experimental data from the $\kappa$ calculated from (\ref{1Bfit}) as follows
$$
b=N^{-1}\sum\limits_{j=1}^{N}\left(\log(\nu(\rho_j))-\kappa\rho_j\right),
$$
where the sum is taken over all $N$ experimental points. 

However, there is no any simple dependence for $b_n$: -5.83, -5.37, -5.09, -4.96, -4.88,
-4.85, -4.84, -4.88, -4.91, -5.03, -5.21 from 1-bromopropane to 1-bromotetradecane correspondingly.

The linear fits determined by the described approximation go through the points calculated from experimental data with the errors, which do not exceed $0.2\%$ for the range from 1-bromopropane to 1-bromoundecane. The deviation starts to be visible for 1-bromododecane and is quite clear for 1-tetradecane. For these two substances, the pairs $(\kappa,b)$, which are equal to $(0.00941,-4.90)$ and $(0.00952,-4.62)$, should be recommended.

Thus, this analysis demonstrates the dependence of the inverse reduced volume fluctuations on chain's length or on the ratio of a number of bromine atoms to the remaining composition of the molecules. This question could be discussed by the comparison with the properties of pure alkanes and alkanes with two substituted bromines considered in the next section.

\section{Discussion and conclusion}

\begin{figure}
\includegraphics[width=\textwidth]{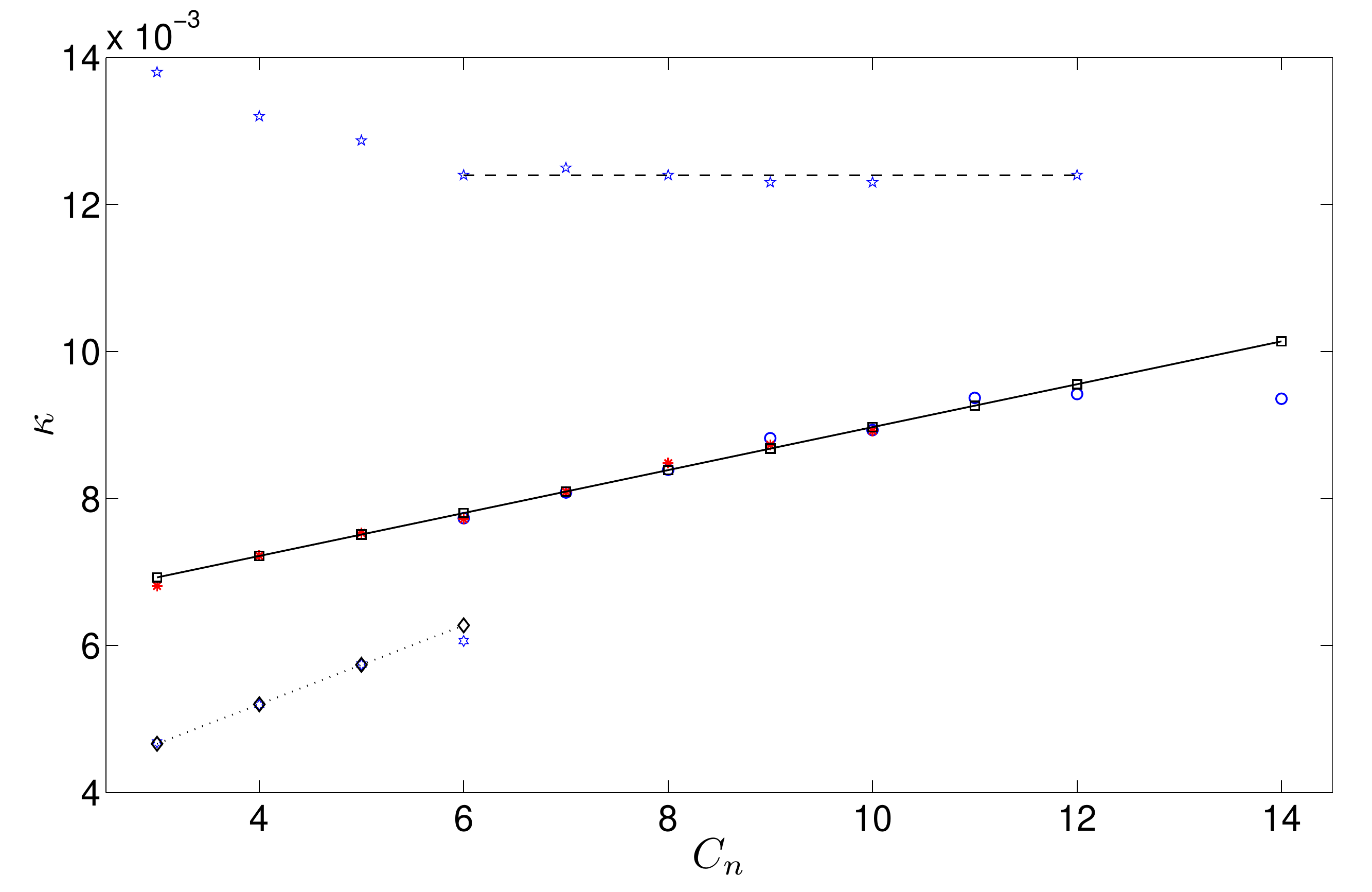}
\caption{The parameter governing a density dependence of the inverse reduced volume fluctuations in 1-bromoalkanes in comparison with n-alkanes and $\alpha,\omega$-dibromoalkanes as a function of the chain's length. The values of $\kappa$ are denoted as follows. 1-bromolakanes: values based on the experimental data presented above (circles) and the data \cite{Ryshkova2009} (asterices), based on the linear fit shown as the solid line (squares); n-alkanes: values from \cite{Postnikov2014} and additionally calculated from NIST data (pentagons) as well as the dashed line $\kappa=0.0124$; $\alpha,\omega$-dibromoalkanes: values from \cite{Chorazewski2015} (hexagons) and based on the linear fit shown as the dotted line (diamonds).}
\label{figallslopes}
\end{figure}

Due to the availability of data for pure n-alkanes and $\alpha,\omega$-dibromoalkanes, we added the plots of $\kappa$ calculated for these sets of substances to those of Fig.~\ref{figallslopes}. The values of the proportionality coefficient between the saturated density and the logarithm of the inverse reduced volume fluctuations for normal alkanes from hexane to dodecane are taken from the work \cite{Postnikov2014}. The rest values (for propane, butane, pentane) are calculated via Eq.~(\ref{nu}) using NIST data \cite{NIST}. One can see that this parameter is decreased for
short-chained n-alkanes, up to pentane. For longer chains, $\kappa$  is practically constant; its value can be taken as $\kappa=0.0124$ providing the practically reasonable accuracy for saturated data. This fact can be explained via the picture of interacting segments of long-chained molecules instead of the whole molecules themselves, as it has been pioneered by Flory \cite{Flory1964}, specially for n-alkanes including the consideration of free volume distributions, which are directly connected with the volume fluctuations. Recently, this scheme is also confirmed by the precious molecular modelling \cite{Zahariev2014}, where the exponential distribution of large holes is detected.

This Flory-like point of view provides a certain insight on the fact of deviation of $\kappa$ for 1-bromododecane and 1-bromotetradecane from the linear fit (\ref{1Bfit}).

It has been shown recently \cite{Chorazewski2015a} that the quantity (\ref{nu}) can be represented for the saturated liquid far below the critical point as 
$$
\nu=\frac{P_i}{T}\left[-\frac{RT}{\mu_0}\frac{d\rho_s}{dT}\right]^{-1},
$$
where $P_i$ is the internal pressure, and the derivative is taken along the coexistence curve.

Thus, the inverse reduced volume fluctuations are determined by both packing properties of molecules and the strength of their intermolecular interactions. The latter is reflected in the value of the internal pressure \cite{Bagley1970,Kartsev2012,Marcus2013}. At the same time, it has been shown \cite{Bolotnikov2009} for 1-bromoalkanes that the averaged energy of intermolecular interaction within the studied liquids can be calculated as $|E|=B\rho^2$ with
$$
B=B_a\left(\frac{\mu_a}{\mu_0}\right)\left(\xi^2+(1-\xi)^2a+2\xi(1-\xi)a^{1/2}\right),
$$
where the index ``a'' denotes the constant of attraction forces and the molecular weight of the reference n-alkanes, $a$ is the ratio between the constant of pair potential of dispersion forces for the halogen and hydrogen atoms, and $\xi=(2n+1)/(2n+3)$ is the share of hydrogen atom centers in the halogenated n-alkane molecule.

As a result, it follows from the equation above that $\xi$ tends to unity for large $n$ and an influence of halogen atoms on the intermolecular energy (and, correspondingly on $P_i$ and $\nu$) diminishes. This line of argumentation explains why the experimental data-based slopes $\kappa$ are quite close for 1-bromoundecane, 1-bromododecane, and 1-bromotetradecane, see Fig.~\ref{figallslopes}, where such a behaviour fulfils for pure n-alkanes, 
 and why the deviation from the hypothetical linear correlation (the solid line in Fig.~\ref{figallslopes}) grows while approaching 1-bromotetradecane. 

On the other hand, the most of available data on~$\kappa$ for $\alpha,\omega$-dibromoalkanes from 1,3-dibromopropane to 1,6-dibromohexane \cite{Chorazewski2015} are also fall into a straight line, see the lower plot in Fig.~\ref{figallslopes}. The equation of this fit is
\begin{equation}
\kappa_n=5.37\cdot10^{-4}n+3.05\cdot10^{-3}.
\label{2Bfit}
\end{equation}

The comparison of linear functions (\ref{1Bfit}) and (\ref{2Bfit}) shows that the slope coefficient for $\alpha,\omega$-dibromoalkanes is practically twice larger than for 1-bromoalkanes. Thus, each bromine atom reduces an intensity of the inverse volume fluctuations
due to the more intensive interaction between them in comparison with
the rest constituents of the alkane chain. The situation is qualitatively the same as in the case of 1-bromoalkanes discussed above. The additional confirmation is the deviation of the point (hexagon), which denotes $k$ for 1,2-dibromohexane from the linear correlation (\ref{2Bfit}):  a longer chain results in the stop of the bromine-dependent growth of $\kappa$. A quantitative estimation of differences between these trends for 1-bromoalkanes and 1,2-dibromoalkanes we leave for further studies, since a certain input may follow
from the intramolecular proximity effect \cite{Kehiaian1983}, which describes a change of the
properties of molecules.

Thus, the study of the inverse reduced fluctuations provides a tool, which allows
for the sequential characterization of the structural characteristics of liquids in their (indirect) interplay with the intermolecular forces. It should be finally pointed out that it is not an abstract problem of molecular physics, but the considered quantity provides an opportunity for accurate predicting thermodynamic values under an elevated pressure \cite{Chorazewski2015,Chorazewski2015a} using the data at normal pressure or saturated data, which can be relatively easily obtained experimentally.

\begin{acknowledgements}
We are grateful to the participants of 11th Winter Workshop on Molecular Acoustics, Relaxation and Calorimetric Methods (03-06.03.2015, Szczyrk, Poland), where the preliminary version of this work has been presented, for fruitful discussions. The work is supported by RFBR, research project No. 16-08-01203А.
\end{acknowledgements}


\end{document}